\documentstyle[vsolj01,graphicx,natbib]{article}

\def\ATel{ATEL}
\def\CBET{CBET}
\def\NewAR{New Astron. Rev.}
\def\Sci{Science}
\def\iaucirc{IAU Circ.}
\def\submitted{submitted}

\def\cite{\citealt}

\begin{document}

\title{Rebrightening Phenomenon in Classical Novae}

\author{Taichi Kato$^1$, Kazuhiro Nakajima$^2$, Hiroyuki Maehara$^3$, Seiichiro Kiyota$^4$}
\author{$^1$ Department of Astronomy, Kyoto University,
       Sakyo-ku, Kyoto 606-8502, Japan}
\email{tkato@kusastro.kyoto-u.ac.jp}
\author{$^2$ Variable Star Observers League in Japan (VSOLJ),
     124 Isatotyo, Teradani, Kumano, Mie 519-4673, Japan}
\author{$^3$ Kwasan and Hida Observatories, Kyoto University, Yamashina,
     Kyoto 607-8471, Japan}
\author{$^4$ VSOLJ, 405-1003 Matsushiro, Tsukuba, Ibaraki 305-0035, Japan}

\begin{abstract}
Two classical novae V1493 Aql and V2362 Cyg were known to exhibit
unprecedented large-amplitude rebrightening during the late stage of
their evolution.  We analyzed common properties in these two light curves.
We show that these unusual light curves are
very well expressed by a combination of power-law decline, omnipresent
in fast novae, and exponential brightening.
We propose a schematic interpretation of the properties common to these
rebrightenings can be a consequence of a shock resulting from a secondary
ejection and its breakout in the optically thick nova winds.
This interpretation has an advantage in explaining the rapid fading
following the rebrightening and the subsequent evolution of the light curve.
The exponential rise might reflect emerging light from the shock front,
analogous to a radiative precursor in a supernova shock breakout.
The consequence of such a shock in the nova wind potentially
explains many kinds of unusual phenomena in novae including early-stage
variations and potentially dust formation.
\end{abstract}

\section{Introduction}

   Classical novae are thermonuclear runaways
(cf. \cite{sta00novareview})
on a mass-accreting white dwarf in cataclysmic variables (CVs).
The light curves of classical novae usually are comprise of rapid premaximum
rise and slower decline from the maximum. 
While many of fast (rapidly fading)
novae show a relatively smooth decline, which is often well approximated
by a power-law (cf. \cite{hac06novadecline}),
slow (slowly fading) novae tend to show more complex behavior.

   In recent years, two fast novae (V1493 Cyg and V2362 Cyg) drastically
violated this picture.  The light curves of these novae initially showed
smooth decline typical for fast novae, but was followed by
accelerating brightening (hereafter rebrightening),
then by a rapid drop (cf. \cite{ven04v1493aqlIRspec}; \cite{kim08v2362cyg}).
The overall feature of the light curves
in these novae was extremely similar
(\cite{gor06v2362cygatel928}; \cite{mun08v2362cyg})
suggesting a common underlying mechanism.

   We here show that the light curves of these two novae can be
very well represented by a combination of power-law decline and
exponential brightening and discuss the origin of the peculiar
light variation.

\section{Data analysis}

\begin{figure}
  \begin{center}
    \includegraphics[width=7cm]{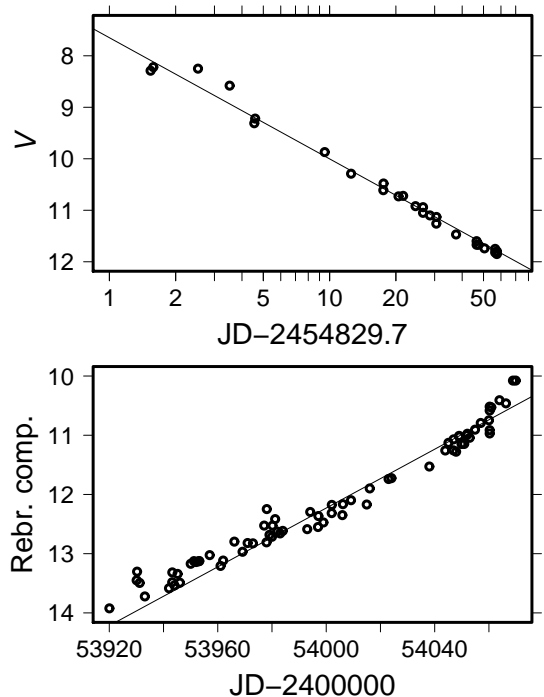}
  \end{center}
  \caption{Best fits to the $V$-band light curve of V2362 Cyg.
  Individual circles represent observations and straight lines
  represent fits with parameters equation \ref{eq:v2362param}.
  (Upper): the early decline is well represented by a power law.
  (Lower): excess rebrightening component (in magnitude scale)
  is well represented by an exponential rise.
  }
  \label{fig:v2362res}
\end{figure}

   The photometric data were taken by us and from observations to VSNET
\citep{VSNET}, supplemented for discovery and early observations
from IAU Circulars.  For V1493 Aql,\footnote{
  The magnitudes at the epoch of the discovery ($m_{\rm pg} = 8.8$ on
  1999 July 13) were \citep{nak99v1493aqliauc7223} later found to
  be incorrect due to the problem in comparison stars
  (vsnet-alert 3254, $<$http://www.kusastro.kyoto-u.ac.jp/vsnet/Mail/alert3000/msg00254.html$>$).  The resultant $t_2$ was $\sim$ 4 d, in contrast
  to the unexpectedly small $t_2 < 3$ d, claimed to be the fastest
  known nova \citep{bon00v1493aql}.
} we adopted visual observations
because CCD observations did not cover the entire stage of the
outburst.  For V2362 Cyg, we adopted CCD $V$ observations.
The typical errors of visual and CCD $V$ observations were 0.1--0.2
and 0.01--0.03 mag, respectively.

   The early part of the light curve is well represented by
a power law-type decline (cf. figure \ref{fig:v2362res}, upper panel),
as is typical for a fast nova.  Subtracting the extrapolated power
law decline, we found that the excess component of the rebrightening
is well expressed by an exponential rise (cf. figure \ref{fig:v2362res},
lower panel).  Upon this knowledge, we modeled the light curve as
a combination (equation \ref{eq:v2362-3}) of two components:
a power-law decline (equation \ref{eq:v2362-1})
and an exponential rebrightening (equation \ref{eq:v2362-2}).

\begin{equation}
  V_{\rm dec} = a + b \log(t-c)
  \label{eq:v2362-1}
\end{equation}

\begin{equation}
  V_{\rm reb} = d (t-t_0) + e
  \label{eq:v2362-2}
\end{equation}
, where $t_0 = 2453920$ is an arbitrary value to avoid large
coefficients.

\begin{equation}
  V = -2.5 \log (10^{-0.4 V_{\rm dec}} + 10^{-0.4 V_{\rm reb}})
  \label{eq:v2362-3}
\end{equation}

  The best-fit parameters were
\begin{eqnarray}
  && a = 7.64 (0.16), \;
  b = 2.35 (0.10), \;
  c = 2453829.7 (0.3), \nonumber \\
  && d = -0.0248 (0.0011), \;
  e = 14.71 (0.17)
  \label{eq:v2362param}
\end{eqnarray}
, for the interval of $2453831 < t(\rm{JD}) < 2454073$.

\begin{figure}
  \begin{center}
    \includegraphics[width=8cm]{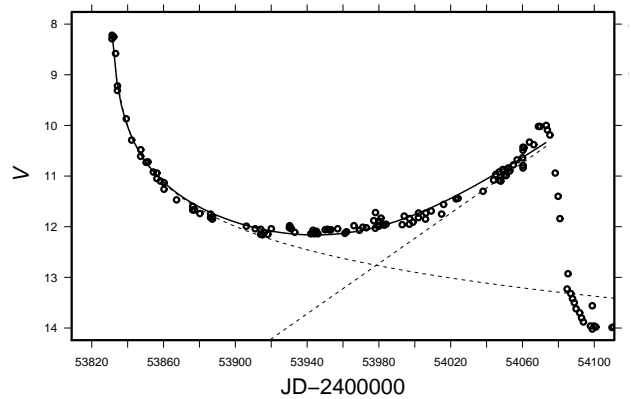}
  \end{center}
  \caption{Best fit of the two-component model (equation \ref{eq:v2362-3})
  to the $V$-band light curve of V2362 Cyg.  The thin dotted curve and line
  represent the components of power-law decline and exponential rise,
  respectively.  The thin curve represents the best fit.
  }
  \label{fig:v2362fit}
\end{figure}

   Figure \ref{fig:v2362fit} shows the overall fit.  The observed light
curve before the rapid fade following the rebrightening maximum is well
expressed by this simple combination of components.
We could not, however, reproduce the result by \citet{kim08v2362cyg}
reporting that an extrapolation of a power-law decline determined from the
first 60 d perfectly fits to the post-rebrightening light curve.
This discrepancy could have originated from the different power indices
between \citet{kim08v2362cyg} and ours. 
In order to estimate the effect of this potential uncertainty in determining
the power index, we calculated fits to different portions
of the early light curve, and a fit to the entire light curve excluding
the rebrightening portion (this simulates the result by \cite{kim08v2362cyg})
and subtracted each of them from the light curve.
The overall trend in unchanged in all cases in that the residual can be
well-expressed by an exponential rise.  The power law index simulating
the result by \citet{kim08v2362cyg} was $b=3.08 (0.04)$, which is
significantly smaller than that ($4.4 = 1.75 \times 2.5$) expected from
the ``universal decline law'' of $F \propto t^{-1.75}$
\citep{hac06novadecline}.

\begin{figure}
  \begin{center}
    \includegraphics[width=8cm]{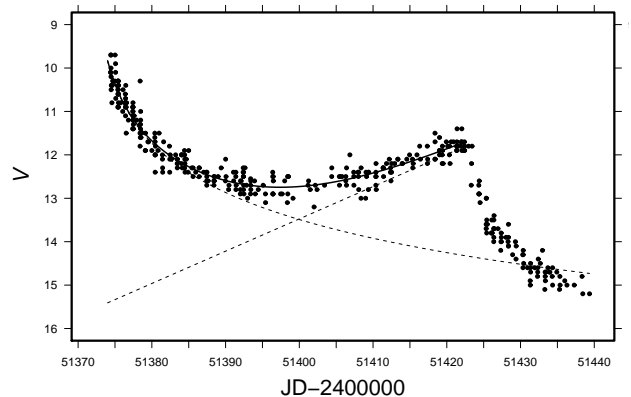}
  \end{center}
  \caption{Best fit of the two-component model to the visual light
  curve of V1493 Aql.  The line are curves are as in figure
  \ref{fig:v2362fit}.
  }
  \label{fig:v1493fit}
\end{figure}

   The same procedure was applied to the light curve of V1493 Aql,
yielding the best fit values (equation \ref{eq:v1493param}) and
the fit (figure \ref{fig:v1493fit}).
\begin{eqnarray}
  && a = 8.75 (0.43), \;
  b = 3.27 (0.37), \;
  c = 2451371.8 (0.6), \nonumber \\
  && d = -0.0744 (0.0008), \;
  e = 13.47 (0.16)
  \label{eq:v1493param}
\end{eqnarray}
, for the interval of $2451372 < t(\rm{JD}) < 2451422$, using
$t_0 = 2451400$.

\section{Discussion}

\subsection{Unusual Rebrightenings and Role of Shocks}

   Several authors have pointed out that the light curves of these two
novae were very unusual but closely resembled each other, but the
the phenomenon still awaits physical interpretation.

   \citet{kim08v2362cyg} tried to reproduce the
rebrightening spike of V2362 Cyg by assuming a temporary expansion of
the pseudo-photosphere of the nova and estimated its temperature and
radius.  \citet{kim08v2362cyg} suggested that this temporary expansion
was caused by a fast ejecta reaching the photosphere, resulting a
significant increase in the density.  Although the authors attributed
the rapid fading following the rebrightening peak to dust obscuration,
the reason is unclear why and how dust formation took place
around the peak.  If this interpretation is also applicable to the extremely
similar nova V1493 Aql, the apparent lack of evidence of dust formation
\citep{ven04v1493aqlIRspec} could be problematic.  Furthermore,
a drastic expansion of the photosphere should affect the free-free
component of the light curve, which is considered to dominate at this
stage of nova evolution (see equation 9 of \cite{hac06novadecline}).
If such an expanded photosphere was obscured by dust formation, and
the remaining light is from a ``detached shell'', the subsequent
evolution of the light curve should follow a steeper decline
(corresponding to the epoch after the wind stops, cf. \cite{hac06novadecline})
rather than a smooth extrapolation of the early decline.
Emission lines would be also affected, while observations of strength
of fitted emission of the main outburst (see figure 4 of
\cite{kim08v2362cyg}) smoothly declined regardless of the rebrightening.
These difficulties probably arise from an assumption of a continuous input
of substantial amount of new ejecta into the preexisting one.
We probably need a more dynamic, rather than this static, process.

   We then propose an alternative idea that the thin shock front formed
by a limited amount of newly ejected matter, and potentially its breakout
in the optically thick nova winds, comprises the rebrightening phenomenon.
The spectra of V2362 Cyg taken around the rebrightening
maximum (\cite{mun08v2362cyg}; \cite{kim08v2362cyg}) showed
temporary appearance of strong and blue-shifted absorption components
in the Balmer-emission-line profiles.  \citet{kim08v2362cyg} also noted
an enhancement of a new component of emission lines during the
rebrightening.  Such high-velocity component is expected to collide
with the exterior, slower, preexisting winds, and will naturally
produce a shock front (cf. \cite{mun07v5558sgrcbet1010}).
The detection of hard X-rays \citep{nes06v2362cygcbet696} can also be
attributed to the presence of a shock \citep{bal98v1974cygROSAT}.
The presence of optically thick wind provides a favorable condition
that a breakout, if present, of such a shock becomes observable
(e.g. \cite{li07sn2006aj} for a supernova/GRB case).
After the optically thick, geometrically thin shock front reaches
the photosphere, and the energy loss via a kinetic motion dominates
over the radiative transport, the subsequent rapid expansion and cooling
of a thin shell can be naturally expected.  The visibility of
the high-velocity component only around the peak indicates that the
velocity structure dramatically varied around the rebrightening peak.
In the present picture, the matter outside the photosphere
would remain largely untouched, preserving the general power law-type
decline trend and behavior of emission lines arising from the outer optically
thin region.  Dust formation may be associated with the quick expansion
of the shock front, as discussed later, probably naturally explaining
the temporal coincidence of the suggested dust formation around
the rebrightening peak.

   The rising stage of the rebrightening, the exponential rise, commonly
observed in these two novae appears to be harder to explain in any
mechanism, and would be a challenge for theoreticians.
Within the present picture, we propose a working hypothesis that 
that photons emerging from a shock front constitute the exponentially
rising, excess component.  Before the shock front reaches the photosphere,
the photons from the shock front have an escape probability of
$e^{-\tau}$, where $\tau$ is the optical depth from the shock front to
the observer.  The observer can thus see photons from the shock front
attenuated by $e^{-\tau}$ as an additional component.
This interpretation corresponds to the ``radiative precursor'' phase
of a shock breakout in a supernova \citep{sch08SNbreakout}.
Under the condition that the front is approaching the photosphere,
$\tau$ is expected to decrease with time.  A simple assumption of
a constant rate of decrease of $\tau$ can explain an exponential rise.
The parameter fits give characteristic time-scales of the rise
corresponding to $\Delta\tau=1$ of 44(1) d and 14.6(2) d for V2362 Cyg
and V1493 Aql, respectively.  The values of $t_2$ being $\sim$9 d
and $\sim$4 d for respective novae, the result might suggest a positive
relation between $t_2$ and the time-scale of the rise.

   Our interpretation is different from that by \citet{kim08v2362cyg}
in that they assumed continuous mass input to the
photosphere during the entire rising stage of the rebrightening
(since day 170, resulting an expansion of the photosphere) while
we assume the continuous presence of a shock front under
the photosphere before the peak of rebrightening
(i.e. the front reaches the photosphere around the time of the peak).
Our interpretation in turn requires the presence of optically thick wind
at the time of the shock formation.  This can be tested
whether the light curve follows by $F \propto t^{-1.75}$ or
by a steeper ($F \propto t^{-3.5}$) power law \citep{hac06novadecline}.
We analyzed the fading part after the end of rebrightening
($2454098 < t < 2454283$).  The best fit powers $p$ for $t^{-p}$ were
1.6(1) for the $V$-band and 1.2(2) for the $y$-band (less affected by
emission lines), which were closer to $p = 1.75$.  This result suggests
that the optical thick winds had not yet stopped at the epoch in question.

   The bluer color around the rebrightening peak, compared to the initial
maximum \citep{mun08v2362cyg} also favors the presece of a hot shock front.
Most recently, \citet{ara09v2362cyg} reported $B-K_s$ multicolor light
curves and attributed the later infrared peak to transient dust formation.
The duration of the infrared peak was unusually short and it quickly decayed
despite its optical thickness.  This makes a contrast to other well-known
novae with substantial dust formation \citep{ClassicalNovae2}.
Although this shortness may be attributed to the only transient formation
of the dust, these light curves, alternatively, can be also interpreted
as progressively slower maxima with longer wavelengths;
the same tendency is inferred from
the optical work by \citet{mun08v2362cyg}.  This characteristic may be
understood as a manifestation of a shock breakout, analogous to
the supernova case (e.g. \cite{kle78SNbreakout}), although a red $V-I$ color
would also require an excess reddening, i.e. the coexisting dust.
This dependence between maxima times and wavelengths would be worth
further investigation.

\subsection{Shock and Dust in Novae}

   A shock formed in an optically thick wind, and its potential subsequent
breakout, is attractive in that it can not only naturally produce
a rapid fade but also enable efficient dust production after
its compression and subsequent rapid expansion in post-shock
cooling as in wind-wind collision binaries (e.g. \cite{uso91wr}).
In V2362 Cyg, at the epoch following the rapid fade,
\citet{ray06v2362cygiauc8788} and \citet{kim08v2362cyg} reported
an infrared excess and reddening which were attributed to dust
formation in the inner region.  The dust formation related to the
expansion after a shock around the contracted photosphere is compatible with
observations only slightly affecting emission lines and the power-law
decline component, which is considered to arise from free-free emission
in outer ejecta \citep{hac06novadecline}.\footnote{
  In V1493 Aql, \citet{ven04v1493aqlIRspec} obtained a spectrum around
  the rebrightening maximum, and detected strong continuum emission
  and low-excitation emission lines.  These findings suggested that
  a second period of continuous mass loss can be responsible for
  the phenomenon \citep{ven04v1493aqlIRspec}.  It is, however, difficult to
  see whether the shock component was present in the spectrum because
  of the low resolution and the very broad emission lines.
  \citet{ven04v1493aqlIRspec} did not report evidence for dust formation.
}

\subsection{Implications to Other Novae and Future Prospects}

\begin{figure}
  \begin{center}
    \includegraphics[width=8.8cm]{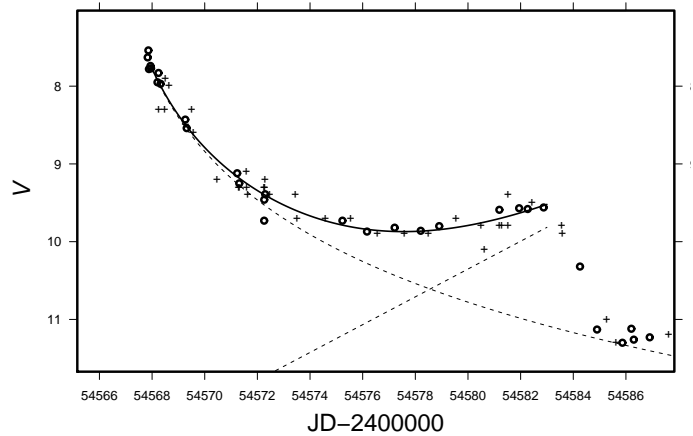}
  \end{center}
  \caption{Fit to the $V$ (circles) and visual (crosses) light
  curve of V2491 Cyg.  The line are curves are as in figure
  \ref{fig:v2362fit}. 
  The parameters were $a = 6.6 (1.2)$,
  $b = 3.6 (1.4)$, $c = -1.1 (1.1)$, $d = 0.18 (0.08)$,
  $e = 11.2 (0.8)$ for the given $t_0 = 2454575$.
  The data were from VSNET.
  }
  \label{fig:v2491fit}
\end{figure}

   We note that similar brightening phenomena accompanied by gradual
brightening and rapid fade, although less conspicuous, are present in
some other novae.  The noteworthy examples include the fast nova
V2491 Cyg 15 d after maximum (cf. figure \ref{fig:v2491fit},
vsnet-alert 10129) and the unusual, slowly evolving nova V1280 Sco
(\cite{das08v1280sco}; \cite{che08v1280sco}),
and potentially V5579 Sgr (vsnet-alert 10201).
\citet{hac09v2491cyg} tried to explain the behavior of V2491 Cyg by
introducing the polar-type magnetism and nova explosion-induced
asynchronization.  Although this interpretation was apparently motivated
by the difficulty in reproducing the rapid decline following the
rebrightening, we do not consider this possibility in the present
paper because the quiescent brightness of V2491 Cyg is incompatible
with a polar and because similar anomalies (high quiescent X-ray
luminosity) were neither detected in V2362 Cyg nor V1493 Aql.

   The fading in V1280 Sco was accompanied by dust formation
occurring at an unexpectedly early epoch.
Although the mechanism of unusual variation in these novae is
not well known, V2362 Cyg/V1493 Aql-type ejection of and shock formation
at an early epoch may have played a role in producing unusual light variation
and in dust production in a harsh environment (cf. \cite{gal77novadust}).
Another commonly seen feature of a bump (maximum) following
a premaximum halt in slow novae (\cite{kat02v463sct} and references
therein) might be an extension of the same phenomenon.
If this is the case, the fading rate (such as $t_2$) derived
from the fading portion of the bump would easily be an overestimate.
The apparent discrepancy between the presence of a premaximum halt and
very rapid fading in V463 Sct \citep{kat02v463sct} might then be solved.

   A model incorporating more elaborate treatment of a shock formation
in realistic nova winds and radiative transfer, which is beyond the
scope of this paper, is needed for a comparison with
observed light curves, color variations, and change in the profile of
emission lines.  Considering the variety of phenomena potentially
arising from this mechanism, a survey in a wide parameter space is
encouraged.
The reason why only some novae show this kind of late-stage rebrightening
is still poorly known.  The present interpretation suggests that if
a second ejection in the late stage somehow occurs when the optically thick
wind is still present, the resultant light curve should generally resemble
those of the two objects investigated.
Such phenomena may have been overlooked in the past
due to the limited high-precision observation.  Future systematic
multiwavelength photometry and high-resolution spectroscopy are desired
to better understand the phenomenon.  X-ray observations are also expected
to provide evidence for the shock in the nova wind.  These observations should
be aimed prospectively starting from the early rising stage of rebrightening.
A deviation of the light curve from the power-law decline, as empirically
shown in this paper, would provide a promising early warning signal for
coordinating such observations, that is, real-time examination of
nova light curves on $\log t-\rm{mag}$ plots will become a powerful
diagnostic tool.

\vskip 3mm

The author is grateful to VSOLJ and VSNET observers who supplied vital data.

\end{document}